\documentclass[prl,twocolumn,showpacs]{revtex4}
\usepackage{bm}
\usepackage{amssymb}
\begin{document}
\title{Black hole singularities: a new critical phenomenon}
\author{Lior M.~Burko}
\affiliation{Department of Physics, University of Utah, Salt Lake
City, Utah 84112}
\date{May 1, 2003}
\begin{abstract}
The singularitiy inside a spherical charged black hole, coupled to a
spherical, massless scalar field is studied numerically. The profile of
the characteristic scalar field was taken to be a power of advanced time
with an exponent $\alpha>0$. A critical
exponent $\alpha_{\rm crit}$ exists. For exponents below the critical one
($\alpha<\alpha_{\rm crit}$) the singularity is a
union of spacelike and null sectors, as is also the case for data with
compact support. For exponents greater than the critical one
($\alpha>\alpha_{\rm crit}$) an all-encompassing, spacelike singularity
evolves, which completely blocks the ``tunnel'' inside the black hole,
preventing the use of the black hole as a portal for hyperspace travel.
\end{abstract}
\maketitle

One of the most intriguing questions in general relativity is the
inevitability of evolution of singularities under very plausible
conditions. Indeed, some view this prediction as evidence that the theory
``carries within itself the 
seeds of its own destruction,'' \cite{bergmann} and look for alternative
theories which are singularity free. An alternative, more tolerant  
approach is to view
singularities as a phenomenon ``from which we can derive much valuable
understanding'' \cite{misner}. Here, we adopt the latter viewpoint. 

There are two ways in which generic singularities in general relativity
are manifested. The first, and the more familiar way, is through unbounded
tidal disruptions, like the singularity inside a Schwarzschild black
hole. Singularities of this type are spacelike and deformationally strong. 
Extended physical objects which approach such a singularity are
unboundedly stretched in one direction and compressed in two other
directions, such that they are inevitably crushed to zero volume. 
The second way in which singularities in general relativity are
manifested is through breakdown of predictability, as with the Cauchy
horizon (CH) singularity inside perturbed spinning black holes. The CH  
singularity is non-central, null, and deformationally weak. That
is, although it is a genuine curvature singularity with infinite tidal
forces, the latter are (twice) integrable, such that the tidal distortions
are bounded \cite{ori}. This singularity signals an inevitable loss of
predictability: to the future of the CH singularity (if the
spacetime manifold can be extended beyond it) lurks a region of spacetime
which is beyond the domain of dependence of any Cauchy hypersurface in the
external universe. The occurrence of the null singularity inside realistic
black holes is of great interest: It leaves open the possibility that
extended objects could traverse the CH only mildly affected,
and reemerge in another universe (or a distant portion of our universe),
in practice using the black hole as a portal for hyperspace travel. 

We consider here the characteristic initial value problem, in which a
pre-existing Reissner-Nordstr\"{o}m (RN) black hole is perturbed
nonlinearly
by a self-gravitating scalar field. The scalar field is specified along
a union of an outgoing and an ingoing null ray, which is our
characteristic hypersurface. 

\begin{figure}
\input epsf
\centerline{ \epsfxsize 7.0cm
\epsfbox{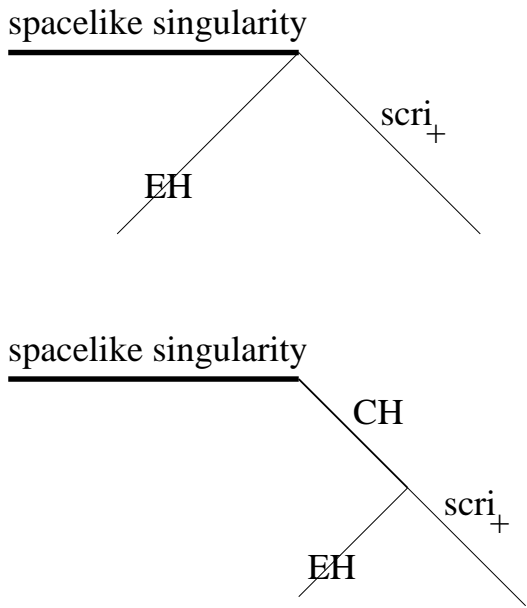}}
\caption{The Penrose diagrams for the two types of singularities. Upper
diagram: an all-encompassing spacelike singularity. Lower diagram: The
singularity includes both a spacelike sector and a null sector at the
CH.} \label{fig0}
\end{figure}

The picture of the singularity inside realistic black holes which has
emerged from analytical and numerical studies which assume
characteristic data with compact support is the following: 
The onset of the null singularity at the CH can be well
described by perturbation theory. The
null generators of the CH focus monotonically, such that the
singular sphere at the CH shrinks as it moves with the speed
of light. Eventually the singular sphere contracts to zero volume, at
which point the causal structure of the singularity changes, and it
becomes spacelike, and deformationally strong. Interestingly, both types
of singularity coexist inside spherical charged black holes, and possibly
also inside spinning black holes. This situation is described by the lower
diagram of Fig.~\ref{fig0}. The spacelike portion of
the singularity inside the black hole is not causally connected with the
null portion of the singularity (except at one point). We can strengthen
(in a sense to be made precise below) the characteristic data of the
(nonlinear) 
perturbation, and thus make the spacelike singularity appear sooner. This
fits well with the general picture: with stronger perturbations the
focusing of the null generators of the CH is faster, such that
the volume inside the contracting singular sphere vanishes 
sooner. 

Can we specify characteristic data which are so strong that a spacelike
singularity evolves {\it before} the onset of the null singularity? Below,
we shall answer this question in the affirmative for a simplified toy
model: Perturbations slightly weaker than a certain critical value have a
singularity described by the above picture, i.e., a union of spacelike and
null sectors. Stronger perturbations involve only a spacelike singularity,
which completely replaces the null singularity and seals off the
``tunnel'' inside the black hole, preventing its use as a portal. 

In this paper we study this question in the context of a spherically
symmetric black hole, which is endowed with electric charge. This is a
simplified toy model for a spinning black hole, as both exhibit similar
causal structures and instabilities of the inner horizons. This black
hole is perturbed by a self-gravitating, spherical, massless scalar
field. Indeed, in this model the above picture was confirmed using both
numerical simulations and analytical analyses for perturbations with 
compact support \cite{brady-smith,burko-prl} (and one family of
characteristic 
data with noncompact support \cite{burko-noncompact}). (The occurrence of
a spacelike singularity inside spinning black holes is yet to be
confirmed.) Any characteristic data with compact support lead to fields
which decay at late times as an inverse power of advanced time along the
event horizon, and lead to the null CH singularity at early
values of retarded times (connected to a spacelike singularity at late
retarded times). Such data are generic, in the sense that they accompany
any realistic gravitational collapse process. The perturbations due to the
collapse can be thought of as a lower bound on the perturbation field. It
is of interest to ask what the singularity structure is when initial data
stronger than the lower bound are present. A simple class of noncompact
characteristic data was studied in Ref.~\cite{burko-noncompact}, where the
profile of the characteristic scalar field was logarithmic in advanced
time. These data preserve the picture of the singularity inside spherical
charged black hole: the singularity is a union of  spacelike and null
sectors. The causal structure of spacetime is again described by the lower
diagram of Fig.~\ref{fig0}. In this paper we show that other families of 
characteristic data may lead to dramatically  different causal structures. 

\begin{figure}
\input epsf
\centerline{ \epsfxsize 7.5cm
\epsfbox{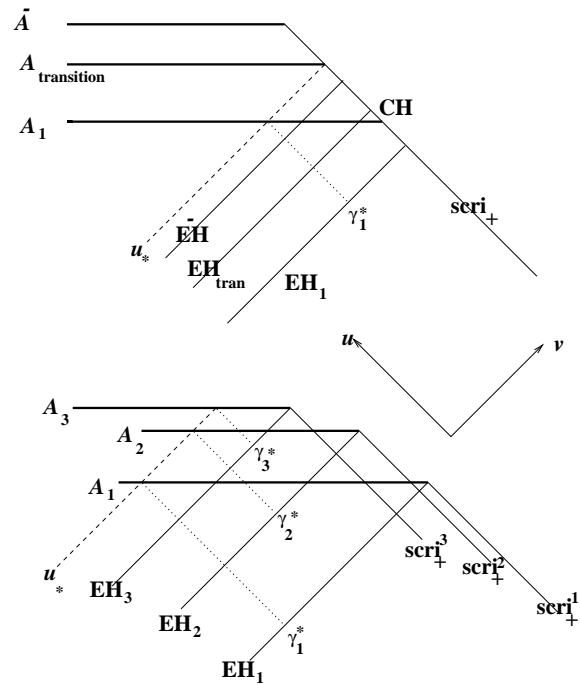}}
\caption{Spacetime diagrams with variable $A$. Lower diagram: Spacetime
diagram for $\alpha>\alpha_{\rm crit}$. Upper diagram: Spacetime
diagram for $\alpha<\alpha_{\rm crit}$. In both cases spacelike
singularities are depicted by a bold solid line. The outgoing null ray
(dashed line) at $u_*$ intersects with the spacelike singularity at a
corresponding affine parameter (along $u_0$) $\gamma^*$ (dotted
lines). With lower value of $A$,the event horizon (EH) moves inward, 
and the spacelike singularity occurs later. }
\label{fig1}
\end{figure}

We write the spherically-symmetric metric in double-null
coordinates in the form
\begin{equation}
\,ds^2=-2e^{2\sigma (u,v)}\,du\,dv+r^2(u,v)\,d\Omega^2 \,
\label{metric}
\end{equation}
where $\,d\Omega^2$ is the line element on the unit two-sphere.
As the source term for the Einstein equations, we take the contributions
of both the scalar field $\Phi$ and the (sourceless) spherical electric
field (see \cite{burko-ori97} for details). The dynamical equations are
the scalar field equation $\square\Phi=0$ and the
Einstein equations, which reduce to
\begin{eqnarray}
\Phi_{,uv}+\frac{1}{r}\left(r_{,u}\Phi_{,v}+r_{,v}\Phi_{,u}\right)=0
\label{KGEQ}
\end{eqnarray}
\begin{eqnarray}
r_{,uv}+\frac{r_{,u}r_{,v}}{r}+\frac{e^{2\sigma}}{2r}\left(1-\frac{Q^{2}}
{r^{2}}\right)=0
\label{EEQ1}
\end{eqnarray}
and
\begin{eqnarray}
\sigma_{,uv}-\frac{r_{,u}r_{,v}}{r^2}-\frac{e^{2\sigma}}{2r^2}
\left(1-2\frac{Q^{2}}{r^{2}}\right)+
\Phi_{,u}\Phi_{,v}=0 \; .
\label{EEQ2}
\end{eqnarray}
These equations are
supplemented by the two constraint equations
\begin{eqnarray}
r_{,uu}-2\sigma_{,u}r_{,u}+r(\Phi_{,u})^{2}=0
\label{con1}
\end{eqnarray}
\begin{eqnarray}
r_{,vv}-2\sigma_{,v}r_{,v}+r(\Phi_{,v})^{2}=0.
\label{con2}
\end{eqnarray}

The initial value problem for these equations and our choice of gauge
for the coordinates are described in Ref.~\cite{burko-noncompact}. We
choose a gauge in which $r=v$ on the outgoing segment of the
characteristic hypersurface $u=u_0$ and $r=r_0+ur_{u0}$ on the
ingoing segment, $v=v_0$. 
We then take the characteristic initial data to have a profile of
$\Phi=A(v^{\alpha}-v_0^{\alpha})$ on $u=u_0$, and $\Phi=0$ on $v=v_0$. The
characteristic value problem is then satisfied by taking $\sigma=\alpha
A^2  (v^{2\alpha}-v_0^{2\alpha})/4-(1/2)\ln 2$ on $u=u_0$ and 
$\sigma=-(1/2)\ln 2$ on $v=v_0$. The parameter $r_{u0}$ is determined by
the initial mass of the black hole $M_{\rm initial}$, its fixed charge
$Q$, and $r(u_0,v_0)$. Here, $u,v$ are (twice the standard) retarded and
advanced times, respectively. Before the onset of the perturbations
($v<v_0$) the geometry is that of a RN black
hole. We then vary the exponent $\alpha$, and study the
occurrence of a null singularity with different values of $\alpha$. 

\begin{figure}
\input epsf
\centerline{ \epsfxsize 8.5cm
\epsfbox{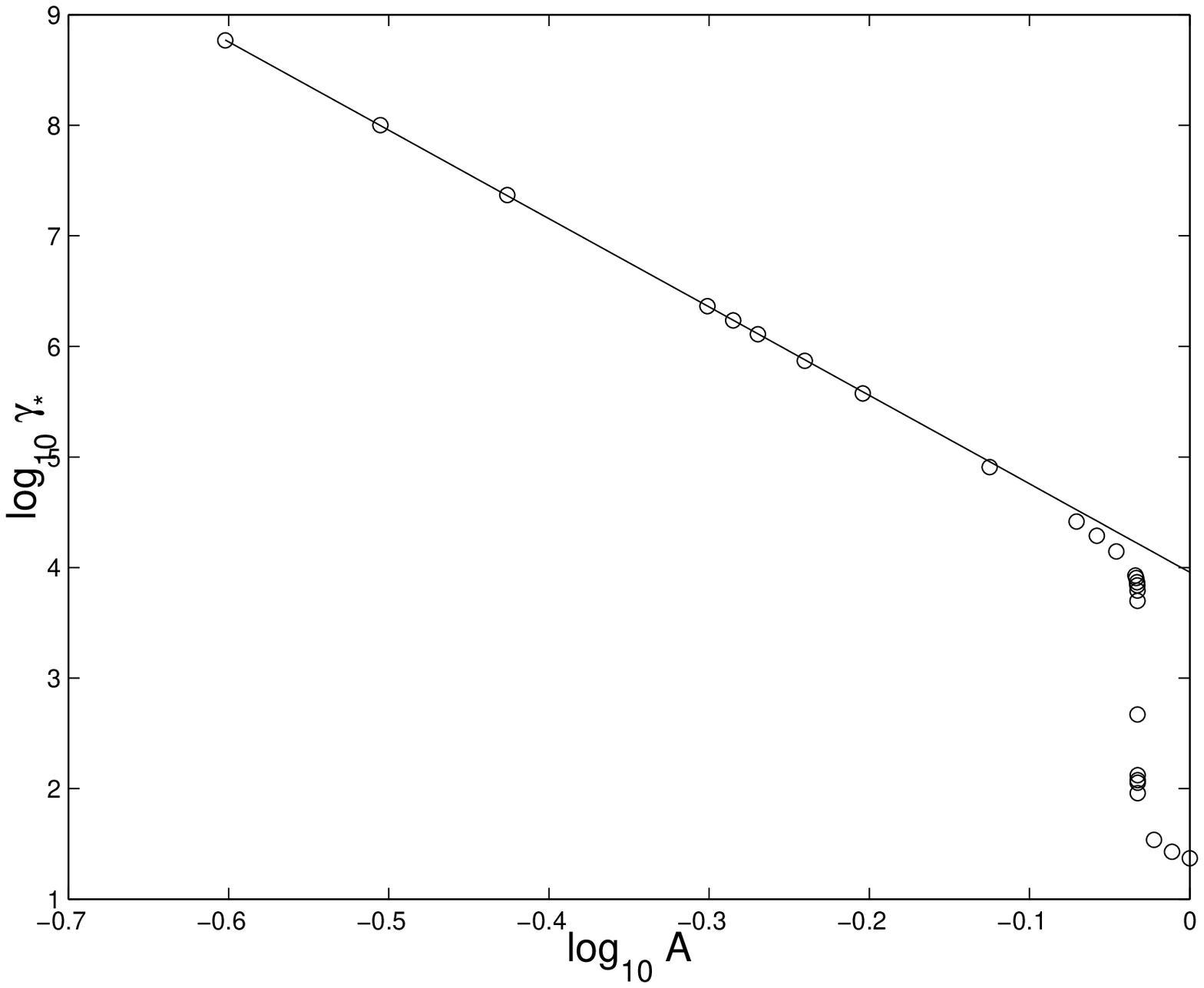}}
\caption{Affine parameter at $u_0$ $\gamma_*$ at which a spacelike
singularity is intersected along an outgoing null ray $u_*$ as a function
of $A$. Here, $\alpha=0.3$.
Numerical data are depicted by circles, and the solid line is
proportional to $A^{-8.00}$.}   
\label{fig2}
\end{figure}

For a given value of $\alpha$, the most straightforward way to establish
the occurrence of a null portion of the singularity is to consider a given
value for $A$, and then probe the fields along various outgoing null
rays. Indeed, that was the method used for finding the null singularity,
and the transition from a null singularity to a spacelike singularity (at
later retarded times) in Ref.~\cite{burko-prl}. In contrast, here we wish
to establish the {\it inexistence} of null singularities for classes of
families
of characteristic data. As numerically it is not easy to locate outgoing 
null rays at arbitrary values of retarded time, and as the level of
required fine tuning is unbounded, we undertake here a
different approach: We fix the retarded-time value of an outgoing null ray
(using the geometry before the onset of the perturbations, i.e., for
$v<v_0$) at $u=u_*>u_0$, and vary the amplitude
$A$ (for fixed values of $\alpha$). (We consider here positive values of
$A$.) That way, we consider a family of spacetimes (sharing a common
$\alpha$). 
In order to compare advanced times between different spacetimes, we
reexpress advanced time in terms of the affine parameter $\gamma$ along
$u=u_0$. It can be readily shown that along $u_0$ (and for fixed angular
coordinates), $\gamma(v)=vM[1/(2\alpha),1+1/(2\alpha),\alpha A^2
v^{2\alpha}/2]$, where $M(a,b,z)$ is Kummer's function. Along $u_*$,
$\gamma(v)$ is of course not an affine parameter, but we can express $v$ 
along $u_*$ in terms of $\gamma$ using ingoing radial null geodesics. We
denote
by $\gamma_*$ the value of $\gamma$ at which a spacelike singularity is
intersected by $u=u_*$. We remark, that these results show that $u=u_0$ is
geodesically complete. Notice that $\gamma(v)$ is a monotonically
increasing function. (We do not have problems using $u$ when
comparing between different spacetimes, because $u$ is defined in terms of
the RN geometry that preceeds the onset of the perturbations.)
With smaller amplitude $A$, the intersection
of the null ray with the singularity occurs at a later value of
$\gamma_*$. (See Fig.~\ref{fig1}.) That is, $\gamma_*$ is a decreasing
function of $A$. If a null singularity exists, there is a finite, nonzero
value $A_{\rm transition}$, such that $\gamma_*\to\infty$ as $A\to A_{\rm
transition}^+\ne 0$. If no portion of the singularity is null,
$\gamma_*\to\infty$ as $A\to 0^+$. [The case $A=0$
corresponds to the exact RN solution (no
perturbations).] That is, the singularity is spacelike for arbitrarily
weak perturbations, and the causal structure of spacetime is
described by the upper diagram in Fig.~\ref{fig0}. 

\begin{figure}
\input epsf
\centerline{ \epsfxsize 8.5cm
\epsfbox{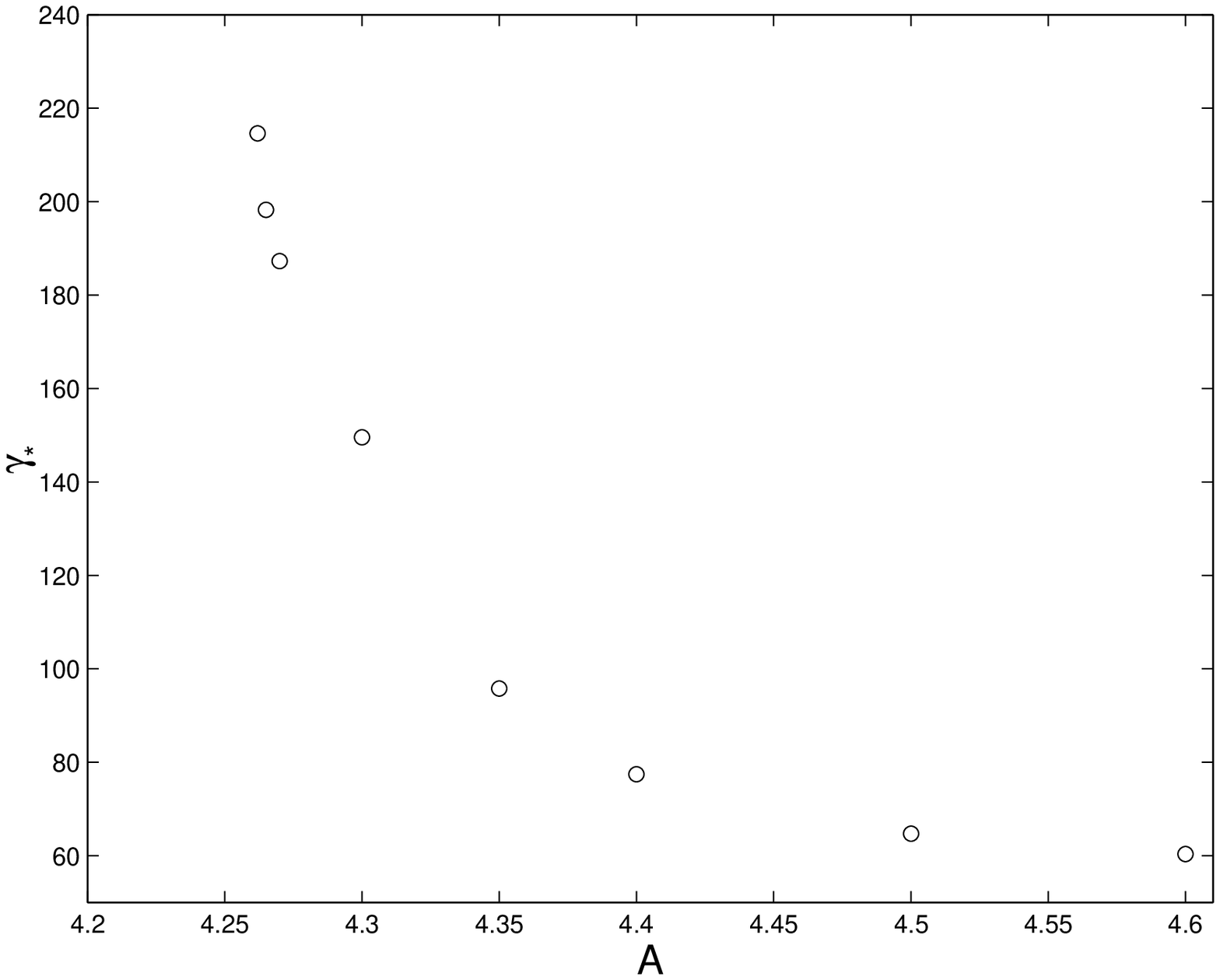}}
\caption{Affine parameter at $u_0$ $\gamma_*$ at which a spacelike
singularity is intersected along an outgoing null ray $u_*$ as a function
of $A$. Here, $\alpha=0.1$. Numerical data are depicted by circles.
}
\label{fig3}
\end{figure}

The results described below refer to the following choice of
parameters: $Q=0.95$, 
$M_{\rm initial}=1$, $r(u_0,v_0)=5$, and $u_*=21.4$. Qualitatively
simlilar results were obtained for other choices of the parameters. 
The numerical code is based
on the code in Ref.~\cite{burko-ori97}. We tested the code and found that
it is stable and second-order convergent. 

Next, we set $\alpha=0.3$. Our results are shown in Fig.~\ref{fig2}, which
displays $\gamma_*$ vs.~$A$. For small values of $A$, $\gamma_*(A)\propto
A^{\beta}$, where $\beta=-8.00\pm 0.01$. Extrapolating this
behavior to smaller values of $A$
we conclude that $\gamma_*(A)\to\infty$ as $A\to 0^+$. That is, for this
choice of $\alpha$ we find that only a spacelike singularity exists, and
that there is no null sector to the singularity. In Ref.~\cite{burko-sl}
it was shown that a spacelike singularity can exist inside spherical
charged black holes. Here, we show that inside some spherical charged
black holes the singularity is only spacelike. 

Figure \ref{fig3} shows analogous results with
$\alpha=0.1$. Here, we find (using Richardson's deferred approach to the
limit) that $\gamma_*(A)\to\infty$ as $A\to A_{\rm transition}^+= 4.26
\pm0.01$. That
is, for $A>A_{\rm transition}$ the outgoing null ray intersects with a
spacelike singularity, and for $A<A_{\rm transition}$ it intersects with a
null singularity. 

In between these two values of $\alpha$ we expect that a
critical value exists, which marks the transition between black holes
whose singularity is composite of two types (spacelike and null), and
black holes whose singularity is an all-encompassing spacelike
singularity. 

\begin{figure}
\input epsf
\centerline{ \epsfxsize 8.5cm
\epsfbox{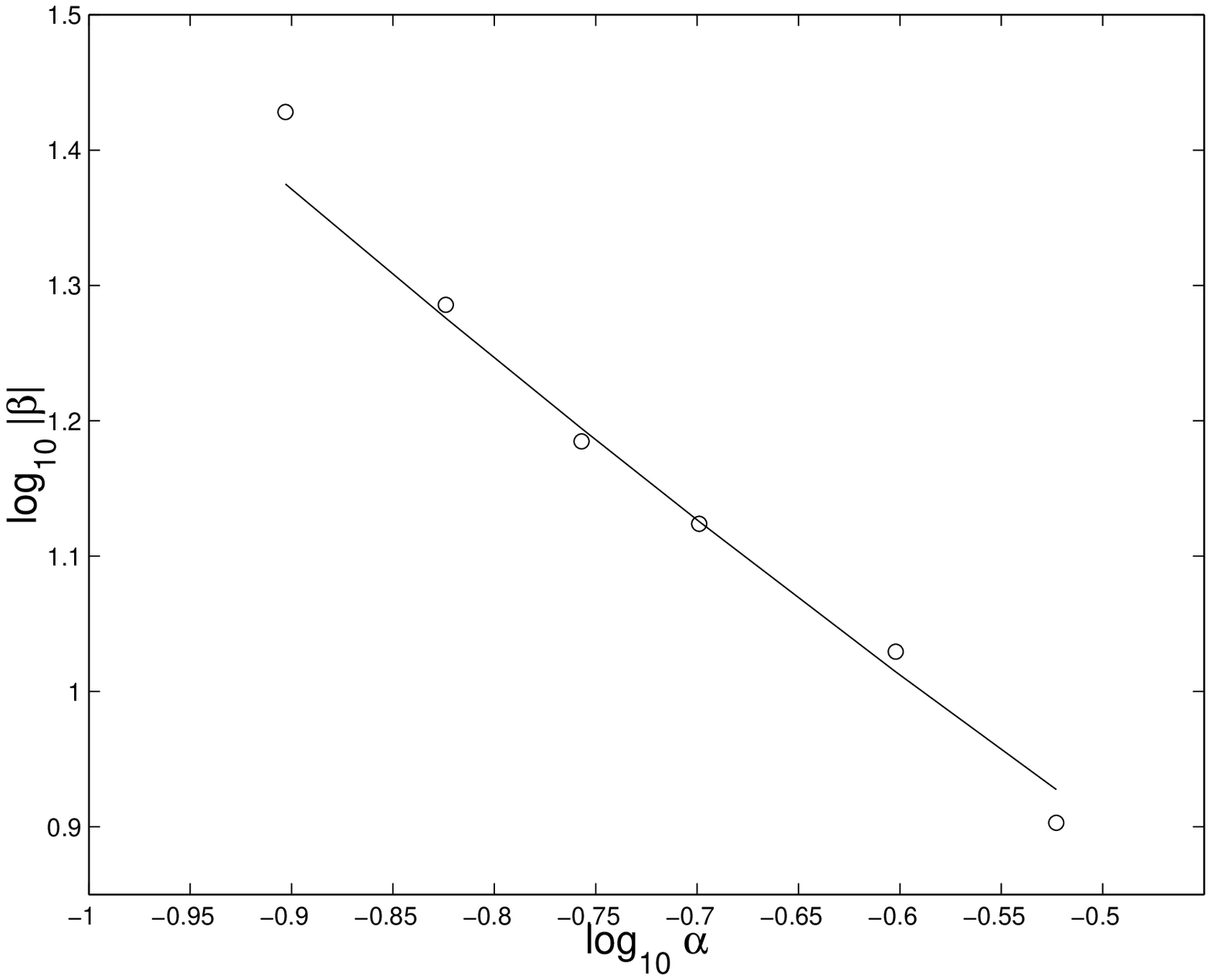}}
\caption{The exponent $\beta$ as a function of $\alpha$. The circles are
the data points, and the solid curve is proportional to
$\alpha^{-1.31}$.}
\label{fig4}
\end{figure}

We return now to the supercritical case in which only a spacelike
singularity exists. The exponent
$\beta$ is a function of $\alpha$. Figure \ref{fig4} shows
$\beta(\alpha)$. The exponent $\beta$ is a monotonically increasing
function of $\alpha$, and approaching (from above) the critical value
$\alpha_{\rm crit}$ its slope becomes steeper. Fitting this curve, we find
that $\beta(\alpha)\propto\alpha^{-1.31\pm 0.07}$. Notice, that $\beta$ is
only defined for $\alpha >\alpha_{\rm crit}$. 

We find that the critical exponent is $\alpha_{\rm crit}=0.12\pm
0.02$. The corresponding value of $\beta$ is $\beta_{\rm crit}=-26.9\pm
3.0$. 
That is, for values of $\alpha$ smaller than $\alpha_{\rm crit}$ we
find that both spacelike and null sectors coexist. For such
values the resulting spacetime has the same causal structure as spacetimes 
which evolve from perturbations with compact support. In particular, the
survival of a null sector leaves open the possibility of harmless
traversal of the null singularity. In contrast, for values of $\alpha$
larger than $\alpha_{\rm crit}$ no null sector survives. The spacelike
singularity is all-encompassing, deformationally strong, and completely
blocks the ``tunnel'' inside the black hole. 

The numerical value we
find for $\alpha_{\rm crit}$ is not expected to be universal, but instead
be model dependent. In particular, we expect it to depend on the ratio 
$Q/M_{\rm initial}$, and the type of field used. In particular, we
expect $\alpha_{\rm crit}$ to increase with $Q/M_{\rm initial}$. Our main
result, namely that there are families of characteristic data for which
the singularity inside a spherical charged black hole is an all-encompassing
spacelike singularity, is expected to be held in general. It is an open
question whether our results hold also for spinning black holes. 
The question of whether the transition we find resembles mathematically
the critical phase transitions of statistical physics (like the threshold
of black hole formation \cite{choptuik}) awaits further investigation. If
so, a possible candidate for a scaling relation could perhaps be the
affine length of the null sector of the singularity as a function of
$\alpha_{\rm crit}-\alpha$. Finally, we comment that realistic black holes
are always irradiated by photons which originate from the relic cosmic
background radiation. This paper then is relevant to the question of
whether these photons completely block the ``tunnel'' inside astrophysical
black holes. The answer to this question certainly depends on the
cosmological model.   

I thank Chris Beetle and Richard Price for discussions. This research was
supported by the National Science Foundation through Grant No.\
PHY-9734871.

\end{document}


\title{Erratum: Black hole singularities: a new critical phenomenon 
\newline [Phys.\ Rev.\ Lett.\ {\bf 90}, 121101 (2003)]
}
\author{Lior M.~Burko}
\maketitle

\setcounter{figure}{2}

The correct expression for the affine parameter $\gamma$ along $u=u_0$ is
given by $\gamma(v)=vM[1/(2\alpha),1+1/(2\alpha),\alpha
A^2v^{2\alpha}/2]$, and not as appears in the original Letter. A number
of quantities which depend on $\gamma$ change as a result of this
correction: For $\alpha=0.3$ the
correct value for $\beta$ is $\beta=-8.00\pm 0.01$, the function
$\beta(\alpha)$ for supercritical evolutions is given by
$\beta(\alpha)\propto\alpha^{-1.31\pm 0.07}$, and $\beta_{\rm
crit}=-26.9\pm 3.0$. The corrected expression for
$\gamma(v)$ also changes Figs. 3,4, and 5. The corrected figures are given
below.

\begin{figure}[h]
\input epsf
\centerline{ \epsfxsize 7.5cm
\epsfbox{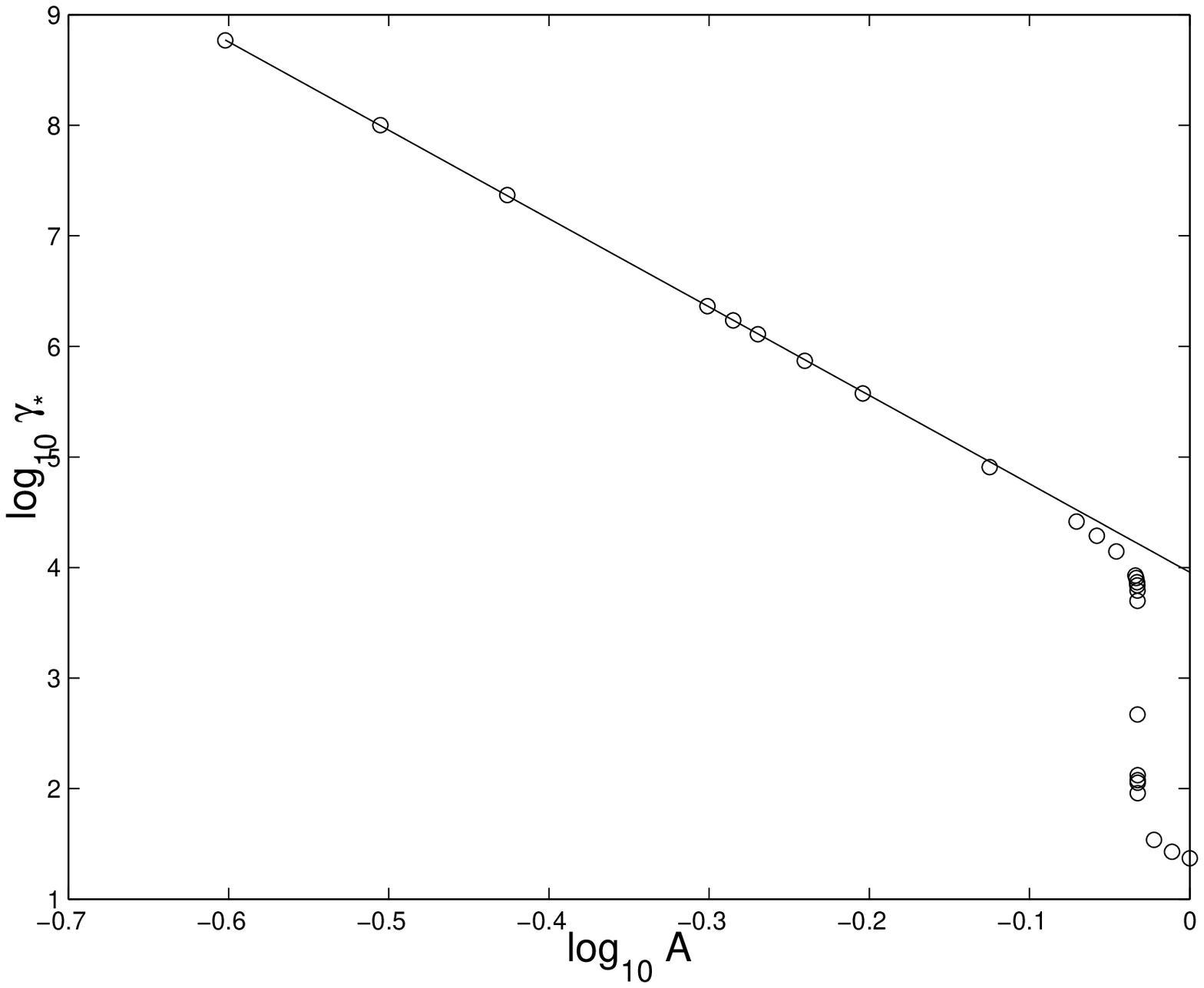}}
\caption{Affine parameter at $u_0$, $\gamma_*$, at which a spacelike
singularity is intersected along an outgoing null ray $u_*$ as a function
of $A$. Here, $\alpha=0.3$.
Numerical data are depicted by circles, and the solid line is
proportional to $A^{-8.00}$.}
\label{fig2}
\end{figure}

\begin{figure}[h]
\input epsf
\centerline{ \epsfxsize 7.5cm
\epsfbox{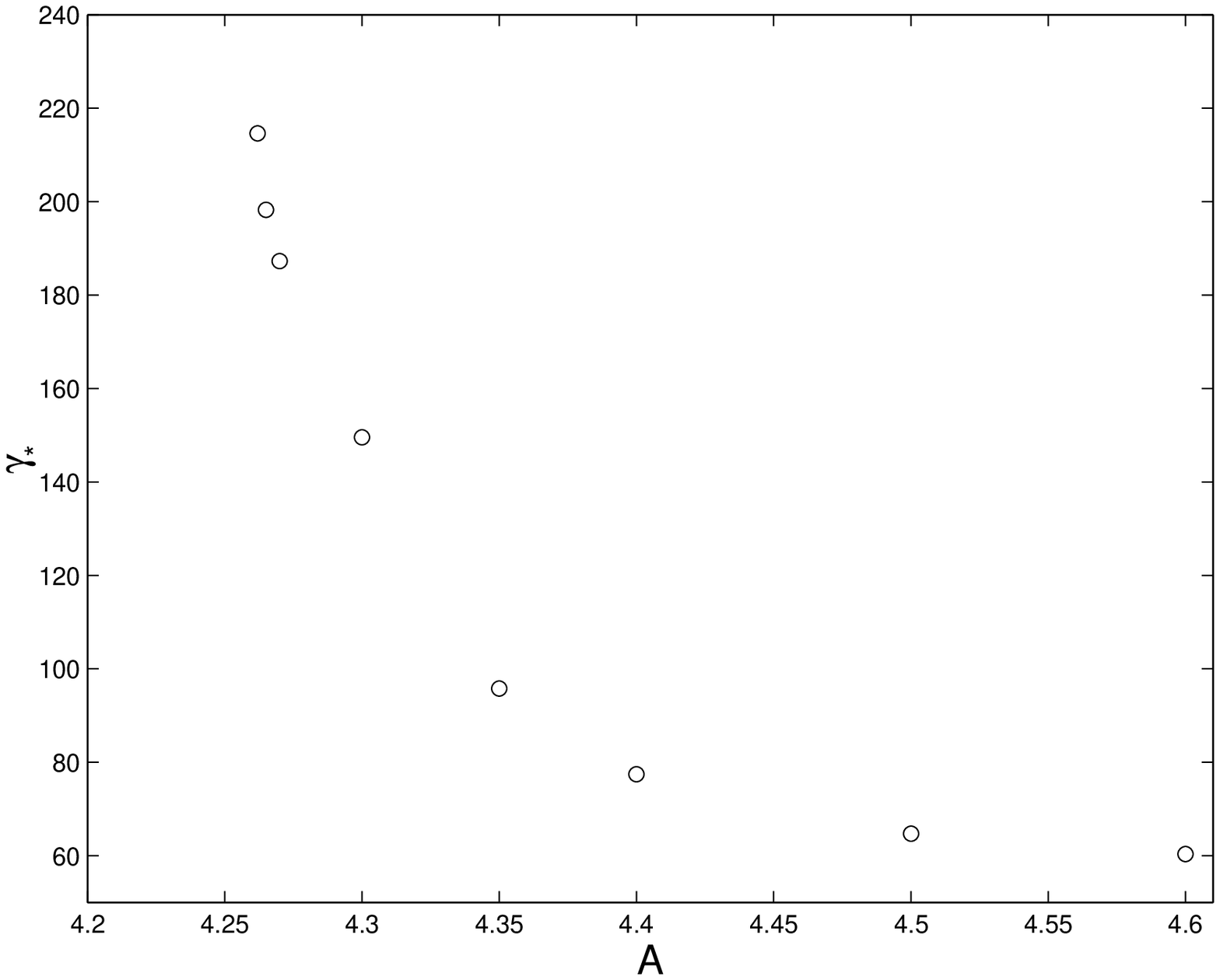}}
\caption{Affine parameter at $u_0$, $\gamma_*$, at which a spacelike
singularity is intersected along an outgoing null ray $u_*$ as a function
of $A$. Here, $\alpha=0.1$. Numerical data are depicted by circles.
}
\label{fig3}
\end{figure}

\begin{figure}[t]
\input epsf
\centerline{ \epsfxsize 7.5cm
\epsfbox{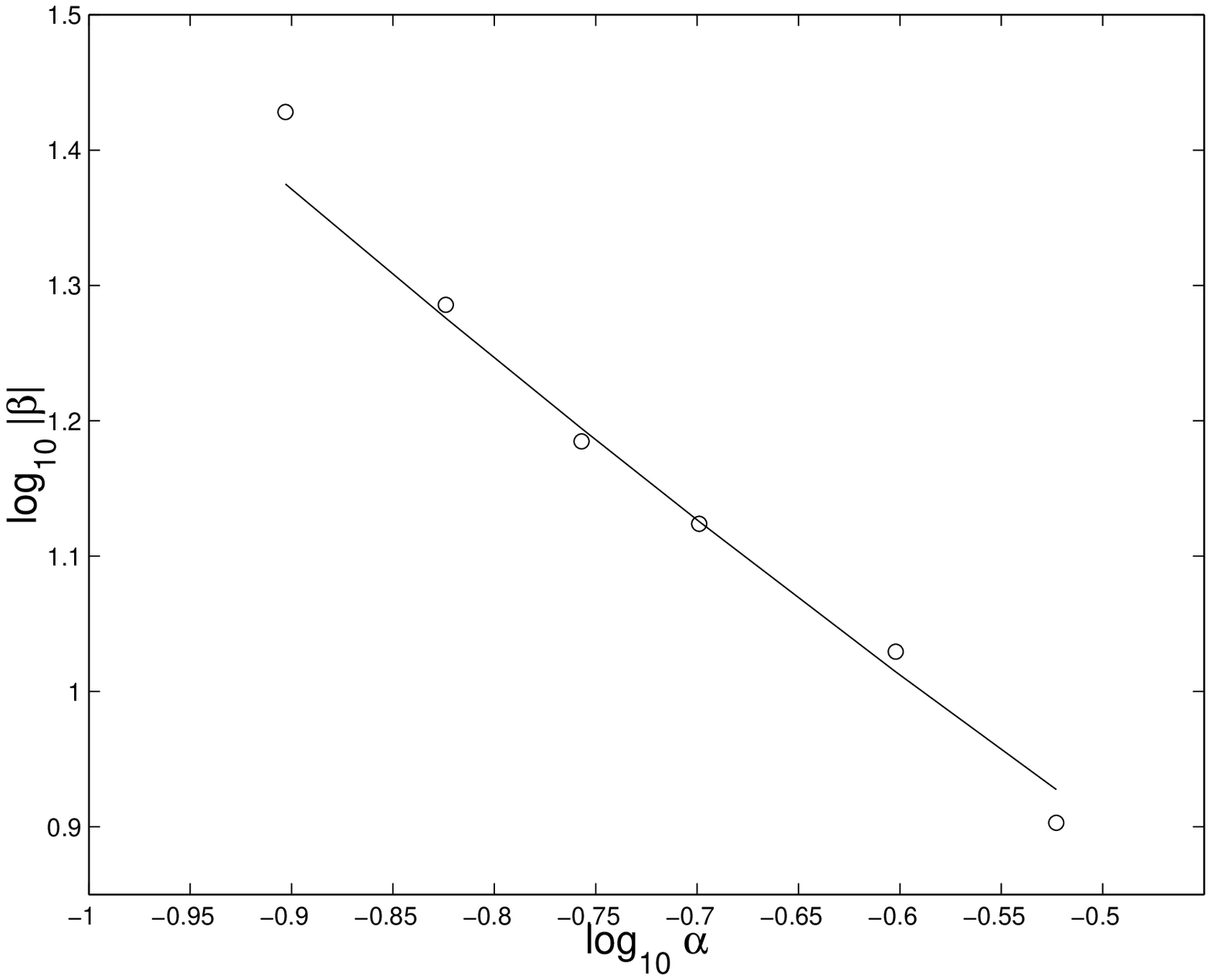}}  
\caption{The exponent $\beta$ as a function of $\alpha$. The circles are
the data points, and the solid curve is proportional to
$\alpha^{-1.31}$.}
\label{fig4}
\end{figure}

As the new function $\gamma$ is a monotonically increasing function of
$v$, and as $\gamma\to\infty$ as $v\to\infty$, it satisfies all the
necessary requirement for comparison between advanced times of different
spacetimes.

The main conclusions of the original paper remain unchanged. In
particular, we find that there are initial data (subcritical data)
for which the singularity
has a hybrid structure (consisting of a spacelike sector and a null
sector), and initial data (supercritical data) for which the singularity
is spacelike throughout.

I thank Aseem Paranjape for pointing out that there was an error in the
function $\gamma(v)$ in the original Letter. This research was supportred
by NSF grant No.\ PHY-9734871.